\documentclass{iau_FM}
\usepackage{graphicx}

\title[Why do hot subdwarf stars pulsate?] 
{Why do hot subdwarf stars pulsate?}

\author[S. Geier]   
{Stephan Geier$^{1,2}$}

\affiliation{$^1$European Southern Observatory, Karl-Schwarzschild-Str.~2, 85748 Garching, Germany \\ email: {\tt sgeier@eso.org} \\[\affilskip]
$^2$Dr. Karl Remeis-Observatory \& ECAP, Astronomical Institute, Friedrich-Alexander University Erlangen-Nuremberg, Sternwartstr. 7, D~96049 Bamberg, Germany}

\pubyear{2015}
\jname{Astronomy in Focus, Volume 1} 
\editors{Piero Benvenuti, ed.}
\begin{document}

\maketitle

\begin{abstract}
Hot subdwarf B stars (sdBs) are the stripped cores of red giants
located at the bluest extension of the horizontal branch. Several different 
kinds of pulsators are found among those stars. The mechanism that drives those 
pulsations is well known and the theoretically predicted instability 
regions for both the short-period p-mode and the long-period g-mode pulsators 
match the observed distributions fairly well. However, it remains unclear 
why only a fraction of the sdB stars pulsate, while stars 
with otherwise very similar parameters do not show pulsations. From an 
observers perspective I review possible candidates for the missing 
parameter that makes sdB stars pulsate or not.

\end{abstract}

\firstsection 

\section{Introduction}             

Hot subdwarfs (sdO/Bs) have been identified as extreme horizontal branch (EHB) stars; i.e. they are core helium-burning stars with very thin hydrogen envelopes and therefore high temperatures. Unlike normal HB stars, which reascend the giant branch as soon as core helium-burning stops, EHB stars evolve directly to the white-dwarf cooling sequence. The formation of hot subdwarf stars in general is still unclear. They can only be formed, if the progenitor loses its envelope almost entirely after passing the red-giant branch. While single-star scenarios are discussed, the focus shifted to binary evolution, when systematic surveys for radial velocity variable stars revealed that more than half of the sdB stars are in binary systems. While the observed close binaries are formed most likely after a common envelope phase, where the companion becomes completely immersed in the red-giant envelopment, stable mass transfer to a main sequence companion and the merger of two helium white dwarfs have been proposed as possible formation channels as well (see Geier \cite{geier15} and references therein).

Several different types of pulsating subdwarfs have been discovered (see Randall et al. \cite{randall14} for a review). The sdBV$_{\rm r}$ stars show p-mode pulsations driven by a $\kappa$-mechanism associated with an enrichment of iron-peak elements in the driving region due to radiative levitation. The sdBV$_{\rm s}$ stars show longer period g-mode pulsations driven by the same mechanism. Hybrid pulsators (sdBV$_{\rm rs}$) showing both kinds of pulsations have been subsequently discovered as well as a variety of pulsating sdOs, He-sdBs and hot subdwarfs in globular clusters. Despite their rather complicated pulsation patterns, hot subdwarfs turned out to be very promising targets for asteroseismic analyses.

Although the different kinds of pulsators occupy characteristic and reasonably well understood instability strips, those strips are not pure. In fact only a fraction of the hot subdwarf stars pulsate, while stars with otherwise very similar parameters do not show pulsations. From an observers perspective I review possible candidates for the missing parameter(s) that make sdB stars pulsate or not.

\section{Abundances}

Radiative levitation is responsible for the enrichment of iron-peak elements in the driving zone for hot subdwarf pulsations. This diffusion effect together with the gravitational settling of light elements can also be observed in the atmospheres of sdB stars, which show characteristic abundances patterns. Several attempts have been made to measure the abundance patterns of pulsating and non-pulsating sdBs from optical and UV spectra and compare them (Heber et al. \cite{heber00}; Blanchette et al. \cite{blanchette08}; Geier \cite{geier13}). However, no significant differences could be found for both the sdBV$_{\rm s}$ and the sdBV$_{\rm r}$ stars and no further conclusions can be drawn.

\section{Stellar winds}

The possible presence of winds in hot subdwarfs has been invoked to explain some of the abundance anomalies observed in these stars, e.g. the helium abundances. Indeed, Unglaub \& Bues (\cite{unglaub01}) showed that the observed helium abundances can be explained if a stellar wind with a mass-loss rate of $10^{-14}-10^{-12}\,M_{\rm \odot}\,{\rm yr^{-1}}$ is present. Vink \& Cassisi (\cite{vink02}) calculated mass-loss rates for EHB stars based on the line-driven wind theory and give upper limits for the mass-loss rates of $10^{-11}\,M_{\rm \odot}\,{\rm yr^{-1}}$. Hu et al. (\cite{hu11}) studied the influence of stellar winds on the pulsation properties of sdBs and found that even very weak winds with mass-loss rates higher than just $\sim10^{-15}\,M_{\rm \odot}\,{\rm yr^{-1}}$ should already prevent the enrichment of iron-peak elements necessary to drive the pulsations.

Stellar wind might therefore be a very promising parameter to explain the presence or absence of pulsations in sdB stars. However, no observational evidence for a stellar wind in a sdB star (e.g. in the line profiles of UV-lines or $H_{\alpha}$) has yet been found. Most recently, we were able to put constraints on the stellar wind of sdB stars using sensitive X-ray observations of sdB binaries with close, compact companions (Mereghetti et al. \cite{mereghetti14}). Wind accretion on the massive white dwarf companion of the sdB+WD binary CD$-$30$^{\rm \circ}$11223 should lead to a detectable flux in soft X-rays even for weak winds. We observed this star with XMM-Newton but could not detect any X-ray emission. However, the derived upper limit of $\sim3\times10^{-13}\,M_{\rm \odot}\,{\rm yr^{-1}}$ for the stellar wind of the sdB is the most stringent observational constraint on this quantity so far. (Pulsating) sdBs with more massive compact companions like neutron stars or black holes, which are actually predicted to exist (e.g. Geier et al. \cite{geier10}) would be very well suited for similar analyses and allow us to either detect the wind accretion or derive more stringent limits.

\section{Rotation}

The excellent, continuous light curves provided by the Kepler mission made it possible to determine the rotational periods of several pulsating sdBs by measuring the splittings of pulsation modes (Pablo et al. \cite{pablo11,pablo12}; Telting et al. \cite{telting12,telting14}; \O stensen et al. \cite{oestensen14}; Foster et al. \cite{foster15}). Preliminary results from long-term ground-based observing campaigns have been published as well (Fontaine et al. \cite{fontaine14}). The derived rotation periods of $7.4-48\,{\rm d}$ are very slow and translate into surface rotation velocities of only $0.2-2.1\,{\rm km\,s^{-1}}$.

We derived projected rotational velocities from the broadening of metal lines in high resolution spectra of 105 sdBs and found that the $v_{\rm rot}\sin{i}$-distribution of this sample is consistent with a uniform rotational velocity of $7-8\,{\rm km\,s^{-1}}$ (Geier \& Heber \cite{geier12}). Although this value is also quite small, it is still considerably higher than the rotational velocities inferred for the pulsating sdBs.

Since the broadening of the spectral lines is not only caused by rotation, the true rotational velocities might be smaller if other yet unknown broadening mechanisms should be present. However, there are no indications for significant micro- or macroturbulence in sdB atmospheres. Furthermore, the measured $v_{\rm rot}\sin{i}$ values and derived angular momenta of the stars are consistent with the ones of the hot blue horizontal branch stars (BHBs) indicating a smooth transition between both populations. Assuming that the rotational velocities derived with both methods are sound, this would mean that pulsating sdBs are rotating slower than non-pulsating sdBs.

To check this, we have to compare the available $v_{\rm rot}\sin{i}$ measurements for pulsating sdBs to the rest of the sample. All our measurements are very close to the detection limit around $5\,{\rm km\,s^{-1}}$, which is given by the intrinsic broadening of the lines and the resolution of the spectra. The small rotational velocities inferred from the mode-splittings of the pulsators are therefore not directly measurable using the alternative method. Only upper limits $<5\,{\rm km\,s^{-1}}$ or marginal detections are expected in this case. Our sample of sdBs with measured $v_{\rm rot}\sin{i}$ contains 13 pulsators. However, eight of them are in close binaries, where tidal effects are expected to influence the rotational velocity (see Geier et al. \cite{geier10}). One of them (KPD\,2109+4401) is a sdBV$_{\rm r}$ with measured RV variations due to strong pulsations. Because the exposure times of the high-resolution spectra necessary to measure the $v_{\rm rot}\sin{i}$ are longer than the pulsation periods, this effect might lead to enhanced line broadening. From the remaining four stars one is in a binary with a period of $10.36\,{\rm d}$, where tidal effects should be negligible.

Three of those stars have $v_{\rm rot}\sin{i}$ measurements close to the detection limit (LB\,1516, $6.0\pm1.3\,{\rm km\,s^{-1}}$; HE\,2151$-$1001, $6.7\pm2.4\,{\rm km\,s^{-1}}$; PG\,1219+534, $5.7\pm1.4\,{\rm km\,s^{-1}}$). The sdBV$_{\rm r}$ star PG\,1219+534 is of special interest, because Fontaine et al. (\cite{fontaine14}) measured a rotation period for this pulsator, which implies a surface rotation velocity of $\sim0.2\,{\rm km\,s^{-1}}$ and is consistent with the marginal detection from spectroscopy. Only PHL\,44 has a higher $v_{\rm rot}\sin{i}=8.4\pm1.0\,{\rm km\,s^{-1}}$. However, based on the data at hand it is too early to conclude that sdB pulsators rotate intrinsically slower than the other sdBs. Several pulsators are found in very close binaries and are significantly spun-up by tidal interactions with the companions. Rotation as such is therefore not preventing sdBs from pulsating. On the other hand, different intrinsic rotation properties might be related to different formation or evolution histories. High-resolution spectroscopy of the Kepler pulsators with measured rotation periods is necessary to resolve this issue. 

\section{Conclusion}

The occurence of pulsations in hot subdwarf stars remains unexplained. In the case of the rather weak g-mode pulsations, the Kepler light curves revealed a higher fraction of pulsators than observed from the ground. Non-detections might therefore be related to limited data quality. However, the fraction of well detectable p-mode pulsators remains small. 

While the surface abundances seem to be unaffected, the strength of the stellar wind might be a crucial parameter. However, current observational limits on the winds of sdB stars are not yet sufficient to test this assumption. The slow rotation measured for the sdB pulsators compared to the other sdB stars might be caused by systematic differences between the methods used. However, it might also be a hint for different formation and evolution histories of both sdB populations.

Most recently, Foster et al. (\cite{foster15}) determined the rotation period of a hybrid sdB pulsator for the first time and detected differential rotation. While the core rotates with a period of $42.6\pm3.4\,{\rm d}$, the envelope, which is probed by the splitting of the p-modes, rotates with a period of only $15.3\pm0.7\,{\rm d}$. Since the rotational broadening measurements of Geier \& Heber (\cite{geier12}) only probe the rotation in the outermost layers of the envelope and the mode-splittings are usually determined from g-modes probing the stellar core (Reed priv. comm.), differential rotation might be a common feature of sdB stars. 

Charpinet et al. (these proceedings) showed that the very slow rotation periods measured for the cores of the sdB pulsators match those of the cores of red giants derived from asteroseismology very well. Since sdBs are regarded as stripped red-giant cores, this makes sense. The differential rotation found by Foster et al. (\cite{foster15}) on the other hand seems to be inconsistent with this result. 

However, the process of stripping the red giant envelope might alter the rotational properties of the core. Hot subdwarfs in close binaries are formed via common envelope ejection and tidal interactions with the close companions lead to a spin-up of the sdB stars. In longer period binaries this spin-up might be more subtle, but sufficient to explain the differential rotation observed. There might even be the possibility that low-mass stellar or substellar companions ejected the envelope, but did not survive this interaction (see Schaffenroth et al. \cite{schaffenroth14} and references therein). Differential rotation might then be an important and maybe the only indicator for such interactions in the past. Although this scenario would not provide an answer to the specific question raised in the title of this paper, it might help to explain the formation of hot subdwarf stars in general.

\end{document}